# Network-aware Adaptation with Real-Time Channel Statistics for Wireless LAN Multimedia Transmissions in the Digital Home

Dilip Krishnaswamy and Shanyu Zhao

*Abstract*— **This paper suggests the use of intelligent network-aware processing agents in wireless local area network drivers to generate metrics for bandwidth estimation based on real-time channel statistics to enable wireless multimedia application adaptation. Various configurations in the wireless digital home are studied and the experimental results with performance variations are presented.**

## I. INTRODUCTION

Streaming multimedia on wireless networks can have high bandwidth requirements, with real-time streaming having stringent delay requirements to be met. Multiple effects including time-varying channel conditions, local or remote congestion conditions, and end-to-end QoS requirements must be matched with an adaptive application capable of offsetting the limitations of the network by managing reliability, latency, and throughput degradations, while hiding the user from the underlying complexity, degraded QoS, and mobility issues in a wireless environment. This paper presents system solutions with network-aware processing agents embedded in WLAN drivers that interact with application-layer software to enable real-time adaptation. Real-time channel statistics monitoring is suggested to monitor link conditions in the network, to enable dynamic multimedia source bandwidth adaptation. Section II provides an overview of different options that exist for proactive adaptation based on dynamically varying wireless link conditions. Section III discusses network-topologies, discusses wireless network statistics that can be monitored and metrics that can be generated with modifications to MAC-level drivers, and how one can accomplish adaptation when such metrics are available. Section IV presents the experimental setup and the algorithms used for the source multimedia node to estimate bandwidth based on the source transmission flow conditions, and based on sniffed real-time wireless link statistics along an end-to-end transmission path. Section V presents results, section VI presents future exploratory options, and section VII concludes the paper.

## II. ADAPTATION OPTIONS IN THE PROTOCOL STACK FOR WIRELESS TRANSMISSIONS

WLANs [1] offer several challenges with regard to multimedia streaming [2, 3, 4, 5, 6]. The effective throughput at the top of the MAC is reduced due to a number of factors [4, 5, 8, 9, 10] such as the number of current users in the network medium, user requirements, priorities, retry-limits, and link adaptation schemes used, channel conditions based on noise and interference, backoff counter depths, backoff stages, protocol timing, and header overheads, and also, the amount of additional time that the medium is unused/idle. The overall throughput is also affected by the transport mechanism used (TCP/UDP/UDP-lite), and by whether there is additional application-layer redundancy such as Forward Error Correction (FEC) across packets being used. In general, the overall throughput as a function of the SINR continues to assume a sigmoidal form [7, 10] with a reduced maximum asymptotic value for the throughput. Application-layer FEC [3, 4] between packets over UDP could be used to compensate for lost packets at the MAC layer, with error-concealment strategies [6] used at a receiver to mitigate the effect of packet losses. One needs to exploit scalability in multimedia representation and identify the most important information to communicate given the available conditions [6]. Application and MAC-PHY cross-layer optimizations [3, 4, 10] and joint source/channel coding [3, 11] can help in adapting to optimally transfer the most relevant information over the wireless channel in response to current channel conditions. MIMO technologies can help in further improving performance [4, 12].

Cross-layer optimizations in ad-hoc wireless networks [13, 14] have been proposed for direct cooperation between layers in a protocol stack to achieve optimal performance. Recent research in the area of streaming wireless multimedia has focused on cross-layer optimizations [3, 4, 5, 10, 15] in the protocol stack at each node. Such optimizations include link adaptation at the MAC layer, retry-limit adaptation at the MAC to compensate for packet errors, application-layer FEC to compensate for packet losses at the MAC, traffic reshaping to handle varying bit rates, dynamic resizing of buffers, management of arrival and departure rates into queues,

Dilip Krishnaswamy was with the Mobility Group at Intel Corporation where this research work was completed. He is now at the Office of the Chief Scientist at Qualcomm, San Diego, USA. His email address is dilip@ieee.org.

Shanyu Zhao is with Digital Home Group, Intel Corporation, 2111 NE 25th, Hillsboro, OR, USA . His email address is shanyu.zhao@intel.com.

reducing end-to-end delay and jitter to meet real-time requirements, adaptive modulation schemes to use more robust modes for base layers, joint source-channel coding, channel reassignment under worsening conditions, and the use of more robust modulation and coding schemes for interference tolerance. A common information base can be used to share information between layers [5]. Channel statistics can be gathered to monitor the usage of the wireless medium so that appropriate decisions can be taken for adaptation. Network-aware processing [16] in a wireless platform is useful for dynamic adaptation based on wireless network conditions. The network-aware intelligent agents used in this paper were implemented at the MAC layer as shown in Figure 1. These network-aware agents interact with the MAC/PHY, monitor the wireless environment, and gather information from the environment to create useful metrics that can be fed back to the application layer to help the application adapt [17]. Here information about available bandwidth and delay/jitter can be used for example, to assist in rate-adaptation at the application layer. Additionally, network-aware agents can also receive information from the application layer that can be used for adaptation at the MAC/PHY, such as taking decisions on retransmission limits or choices for modulation and coding schemes to use based on packet priorities; however such MAC/PHY adaptation is not the focus of this paper. This paper focuses on cross-layer adaptation where channel statistics from the MAC/PHY are obtained to create useful metrics to enable the application to adapt. The information content in a video stream can be used to exploit scalability in different forms such as spatial, temporal and SNR scalability. When bandwidth fluctuations occur in a wireless LAN environment, a multimedia application can be designed to gracefully adapt to such changing conditions by monitoring changes in the channel conditions, and by exploiting the scalability inherent in the multimedia representation.

III. NETWORK-AWARE INTELLIGENT AGENTS FOR ADAPTIVE MULTIMEDIA PROCESSING

Consider a multimedia server providing a video stream (such as an MPEG2 transport stream) to a client in a wireless LAN network. It is possible that there may be other streams occupying the same channel in the network at the same time. Some of the other streams could be carrying multimedia as well. The objective is to determine the available bandwidth in the network, and for a multimedia streaming server to adapt to the dynamically changing conditions in the network. Streaming could occur over multihop 802.11s mesh networks as well. Real-time channel statistics under varying channel conditions can provide faster and more accurate feedback for cross-layer optimization. Typical monitored statistics can include transmit time, backoff time, other used time, and idle time (see Figure 2). From the idle time, and the current partitioning of time between transmit time and backoff time, one can determine how much of idle time can be used for transmitting. A knowledge of the transmitted bytes within the transmit time (which is a function of the modulation and coding schemes used) will provide information to the multimedia server about how many bits/second it can send in the next time window.

The backoff time selected randomly for each transmission needs to be monitored. For one-way high bandwidth transmissions, TX statistics are most important. For two-way transmissions, both TX and RX statistics are important. For Delay/Jitter info, knowledge of the transmit queue depths for each of the access categories is important. Knowing additional information such as the number of burst errors and total number of packet errors in a window can help in additional optimization such as application layer FEC when using UDP as the transport protocol. Time utilization with respect to a node in a wireless LAN network can, in general, be broken down into several components.

*Total time = LocalTransmissionTime + LocalBackoffTime + OtherUsersTransmissions + NetworkIdleTime ……………(1)*

Given the CSMA/CA nature of transmissions in a WLAN network, available time on the network needs to be shared between different users active in the network. In addition if there is other unrelated traffic in the network, bandwidth needs to be shared with those traffic components as well. Some amount of time is spent by the node in wireless transmissions including protocol timing overheads, and time spent in receiving an acknowledgement. This contributes to the *LocalTransmissionTime*. A certain amount of time needs to be allocated for the random backoff counter prior to transmission, and this is accounted in the *LocalBackoffTime*. Some additional time may be used by other users for their transmissions including protocol timing overheads and acknowledgements. This can be absorbed into the time allocated for *OtherUsersTransmissions*. For a saturated network with several users, the *NetworkIdleTime* can be assumed to tend to 0; however it is possible that a significant non-zero idle time is perceived in a saturated network, when no users are transmitting because all users may be at the backoff stage and counting down on their backoff counters.

Within a period of monitoring time *measTime*, one can measure the transmission time *(TxTime)* and the backoff time *(BackoffTime)* associated with all multimedia packets (number of bits transmitted = *TxBits*) at the PHY/MAC layer in the protocol stack. The time *TxTime* refers to both the actual transmission and the protocol overhead times. In a WLAN network with 802.11e priority queues enabled [18], one has to monitor this information with respect to each queue. The effective throughput for the multimedia traffic is given by *TxBits/measTime*. Here *idleTime* refers to the time when the local node is neither transmitting nor receiving nor is it in a backoff stage, and nor is any other node transmitting over the medium. Therefore *idleTime* refers to the network idle time as perceived at a given node with respect to the wireless medium. Now, if there is a part of *idleTime* available during *measTime*, then such additional time could have been used for increasing the bandwidth associated with the multimedia traffic. Since the statistics measured may not be completely accurate, and other users may join the network or attempt to increase their data rates, it may be wise to be conservatively opportunistic and request for only a fraction $\rho$ of the available *idleTime* on the network. Since *TxBits* were transmitted in the time *(TxTime + BackoffTime),* then one can expect that *TxBits * ρ * [idleTime / (TxTime + BackoffTime) ]* would be the additional bits that can get transmitted in the additional time available. This assumes that the relative distribution of time between *TxTime* and *BackoffTime* remains unchanged.

When a multimedia stream has to traverse multiple hops, the available *idleTime* needs to be shared across the hops, and hence one may have to choose a smaller $\rho$ to allocate sufficient time for data transfers on each hop. In general, with the availability of the interference graph for the mesh network, and based on the channel allocation to links in the network, one can identify the links on the same channel that contend for the network. As link information about multiple hops arrive using cross-overlays, the multimedia source nodes can be expected to adapt with better [19, 21]. Incremental rate adaptation can also be employed where rate increases are conservative with a fraction of the predicted available bandwidth [17]. As shown in the algorithm in Figure 3, rate adaptation feedback can be positive (factor $\rho$) or negative (factor $\beta$) based on the current channel conditions. As additional statistics arrive for subsequent time intervals, further incremental rate adaptation can be attempted to encroach on the remaining available idle time, until the rate adaptation converges.

## IV. EXPERIMENTAL SETUP AND ALGORITHMS USED

The following nomenclature is used to describe the topologies considered for experimentation: *g* represents an 802.11g wireless link, *a* represents an 802.11a wireless link, *AP* represents an Access Point, w represents a wired link, *dls* means Direct Link Setup, *Xn* represents a cross-traffic of *n* Mbps imposed (in addition to any environmental congestion, interference, and noise). Based on the above nomenclature, Figure 4 shows all topologies used in our experiments: The metrics used for adaptation need to be measured in the wireless MAC driver and accumulated over a measurement time interval (such as a 200ms interval). The measured metrics can be propagated to the application layer for optimization.

### A. Bandwidth estimation metrics based on real-time statistics

**Source Predictor (SP) estimation** is used when a multimedia source is transmitting wireless data over the first hop. The multimedia source observes transmission statistics over a measurement interval of *M* ms, and then adapts the application data rate if required. For a multi-hop configuration where the same channel is used to transmit bits over additional hops, a correction factor "*p*" is used to account for the need to share the idle time over all hops using the same channel.

SP Estimated Additional Bandwidth = $\frac{I}{M} \cdot \frac{TxBits}{Tx + Bo} \cdot p$ ..(2)

In the above formula, *I* is idle time, *M* is measurement time, *TxBits* is the transmitted bits during the measurement time, *Tx* is the transmission time, and *Bo* is the back off time. It should be noted that the *Tx* includes time for preamble, inter-frame space, RTS/CTS, and ACK packets. This bandwidth estimation technique therefore uses the sample of packets transmitted in a given measurement interval to calculate the bit-rate. It assumes that if the sender sents out another *TxBits* bits in the same time slot, it would have used *Tx* time for sending and *Bo* time for backing off. Given the idle time available relative to the measurement time interval, one can then estimate the additional bandwidth available assuming a proportional distribution of the available idle time for transmitting additional data. The factor "*p*" is a proportionality factor that is used to correct the possibility of overestimation of idle time as perceived in the network by the MAC driver. Typical values of *p* used are 0.8 for a single hop and 0.4 for a two-hop configuration. However, these values can vary depending on the relative qualities of the links in each hop. SP estimation suffers from lack of information of additional hops in a path, if only statistics are available about the first link connected to the video source. However, with cross-overlay feedback, more accurate relative utilization of the links can be obtained based on link qualities.

**Source Sniffer (SS) estimation** can be used if the network-aware multimedia source sniffs and measures the quality of all the wireless links in the transmission path. With the knowledge of the available idle time in the network, the available bandwidth is estimated as follows:

SS Estimated Additional Bandwidth = $\frac{I}{M} \cdot \frac{1}{\sum(\frac{r_i}{q_i})} \cdot f$ ...(3)

In this formula, *I* is channel idle time, *M* is measurement time, $r_i$ is average retry rate of link *i*, and $q_i$ is the average physical layer transmission rate of link *i*. *f* is a correction factor that accounts for the throughput drop that one observes at the top of the MAC layer relative to the physical layer transmission bandwidth based on the modulation and coding scheme chosen. For example, if the 64QAM, 3/4 code rate modulation and coding scheme is chosen for transmission over an 802.11a link, then the physical layer rate is 54Mbps, while the corresponding MAC-layer effective bandwidth is likely to be about 24Mbps (it can be lower depending on the congestion in the network and additional overheads in the network). In this example, *f* can be chosen to be 24/54 = 0.444. With only one transmission attempt at phy-rate *q* and *r = 1*, the bandwidth becomes *( I/M) * q * f*. This then translates to *f* being a fraction accounting for performance losses in the protocol stack relative to the physical layer rate available. The retry rate is defined as the ratio of the number of transmitted attempts of all packets intended for transmission including all retransmission attempts of the packets, divided by the number of packets intended for transmission.

## V. EXPERIMENTAL RESULTS

For each experiment setup, we first measure the capacity (the maximum bandwidth) from the source to the destination, by pushing as much data as possible without limiting data rate. Then we send a trans-rated version of MPEG2 video streams (UDP or TCP traffic) from source to destination at a certain data rate for 30 seconds, e.g., 1Mbps, 2Mbps, 4Mbps, etc. The available bandwidth (SP and SS metrics) was calculated based on the statistics available from the driver. We then show that the actual data rate plus the available bandwidth should approach the capacity of the path. We have two types of graphs, Per-Measurement graph showing "actual bandwidth + available bandwidth" for each SP/SS statistics (about 200ms), and a summary graph showing "average actual bandwidth + average available bandwidth" for the whole measurement period (30 seconds). The experiments were conducted both in a Screen Room (a room with copper screens for isolating outside wireless signals) and in an actual house in a farmland.

*Per-Measurement graphs*

Per-Measurement graphs show the estimated available bandwidth for each measurement timeslot, generally 200ms. Figures 5 and 6 show results form experiments in the Screen Room. It can be seen that the total estimated bandwidth (measured bandwidth + estimated available bandwidth) tracks the capacity in the channel quite well, with both SP and SS bandwidth estimation techniques. The dark orange line in the middle represents the capacity and the two green lines around it demarcate the 20% variation area. It should be noted that dynamic fluctuations in conditions in the channel can cause the actual estimated capacity to vary relative to the average channel capacity.

### A. Summary graphs

Figures 7 and 8 show SS vs SP comparisons for one-hop and two hop bandwidth estimation in the Screen Room. Figures 9 and 10 show similar comparison for two-hop configuration in a Farm House. Figures 11 and 12 show reduction in available bandwidth in the presence of a 5Mbps cross-traffic stream. Both SP and SS estimation algorithms work reasonably well in the presence of cross traffic. In general, there are times when the available bandwidth is over-estimated, and at other times the available bandwidth is underestimated. Parameters such as *f* and *p* that were used are only approximations. In addition, idle time measurements can include hardware delays when packet buffers may be empty with variable bit rate traffic. This results in an underutilized channel which can only be corrected if traffic is shaped for better channel utilization. When the channel saturates, idle time measurements may be non-zero but unvarying even if additional data is pumped into the network. Such a situation indicates that there is no additional bandwidth available in the network. When the channel is unsaturated but conditions deteriorate, the transmission times may go up, but the number of *TxBits* transmitted may continue to match the expected approximate source transmission rate. However, when the channel is saturated and channel conditions deteriorate, then *TxBits* reduces, which will trigger a reduction in the multimedia source rate at the application layer (Figure 3). At the application layer, a correction factor needs to be imposed to reflect the additional bits accumulated by application layer data as it traverses down the protocol stack for transmission. At the video application layer, incremental adaptation was used with a conservative version of the estimated available bandwidth being used to increase the application throughput. As successive measurements were obtained the application-layer throughput was refined. This was to prevent the application from assuming that there was more throughput than was actually available which would have led to frame losses, resulting in lower video quality as estimated by the Video Quality Metric (VQM) tool [20]. Improvements in video quality were observed using this subjective tool when the feedback from the network-aware agent was used for multimedia bandwidth adaptation. However, PSNR calculations were not done to compare the difference in video quality for the current set of experiments.

## VI. FUTURE EXTENSIONS

Future extensions could include more accurate dynamic network-aware information to adapt the multimedia streams. For general wireless topologies such as a multi-hop mesh network topology, it is necessary to have information transfer between nodes in the network, so that the nodes can become aware of link conditions in the entire network. A generalized network-aware architecture [21] [22] would help in scaling the solution further. Cross-layer interactions can be used to exchange information between layers. A cross-overlay architecture can be used to exchange information using distributed virtual machines in the application layer. Such architecture has been demonstrated for distributed routing optimization for multiple flows for VoIP applications with end-to-end delay estimation in [21] with a pre-standard 802.11s implementation at the link layer. The architecture needs to be extended to support video applications as well, for end-to-end delay and end-to-end bandwidth estimation. In a multi-hop path, one needs more accurate information about links further away from a source. Sniffing link traffic to estimate link qualities at just the source node (SS estimation) can only help when transmissions are within range of the source. Corrective fractional factors to account for loss in throughput due to sharing of channels across multiple hops with link information available only at the source node (SP) can at best be approximate. As link information about multiple hops arrive using virtual machines to exchange cross-overlay information, the multimedia source nodes can be expected to adapt with better knowledge about network conditions.

## VII. CONCLUSIONS

This paper has suggested practical algorithms to be employed in intelligent agents for network-aware wireless multimedia adaptation. Real-time wireless statistics were monitored, and useful metrics were generated for bandwidth estimation for proactive wireless multimedia adaptation in the wireless digital home. Different techniques were explored for adaptation and experimental results were presented for various network and traffic configurations. The monitored values of the idle time were used to estimate the available network bandwidth which appeared to be quite close to expected values. The improvement in video quality based on the feedback from the network-agents was monitored by a subjective video quality measurement tool such as VQM and improvements in video quality were observed. In general, a conservative and graceful approach to adapting to rate changes proved fruitful in the wireless digital home. Future solutions would include extending the current techniques to multi-hop mesh scenarios.

VIII. ACKNOWLEDGEMENTS

Mousumi Hazra from the Mobility Group at Intel has provided many thoughtful ideas and feedback to this paper. Rik Logan at the Intel Digital Home Group facilitated good discussions, and equipment and venues for the experiments. This work could not have been accomplished without their help.

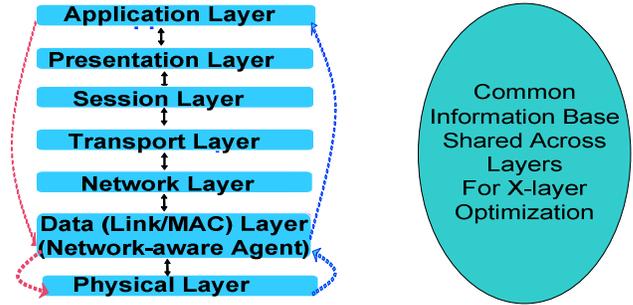

**Figure 1: Network-Aware Cross-Layer Optimization**

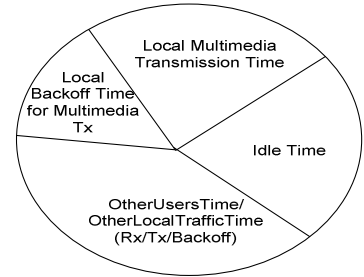

**Figure 2: Pie Chart of Time Utilization**

*TxRate = TxBits/measTime ;*
*ΔTxBits = 0;*
*If (IdleTime>IdleMinThreshold)*
    *ΔTxBits = ρ * TxBits * [IdleTime / (TxTime + BackoffTime*
            *+ PacketProcDelay ) ];*
*else*
    *TxRateDiff = TxRatePrevious – TxRate;*
    *If (TxRateDiff > MinRateDiffThreshold)*
        *ΔTxBits = - β * MeasTime * TxRateDiff;*
*ΔTxRate = ΔTxBits/MeasTime;*
*AvailTxRate = TxRate + ΔTxRate;*
*TxRatePrevious = TxRate;*
*TxDelayPrev = TxDelay;*
*TxDelay = TxQueueDepthInBits/TxRate;*
*TxJitter = TxDelay – TxDelayPrev;*
*ReturnToAppLayer(TxRate,AvailTxRate, TxDelay, TxJitter);*

**Figure 3 : Incremental Rate Adaptation Algorithm**

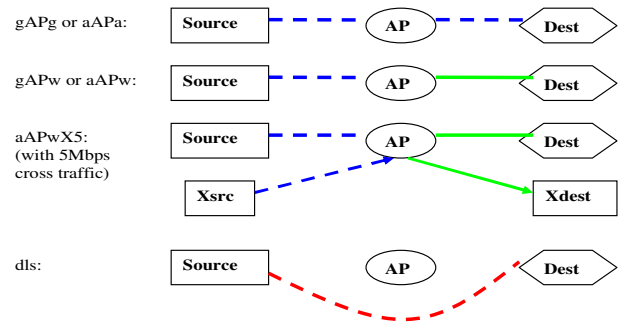

**Figure 4: Wireless multimedia transmission topologies considered for experimentation**

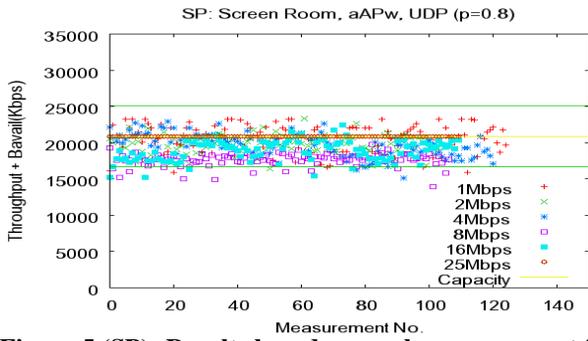

**Figure 5 (SP): Results based on each measurement for SP estimated bandwidth for single hop wireless UDP traffic.**

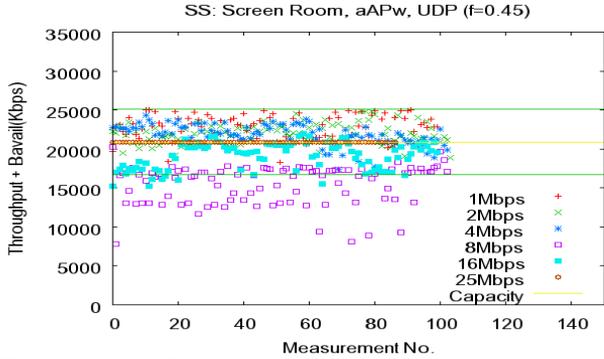

**Figure 6 (SS): Results based on each measurement for SS estimated bandwidth for single hop wireless UDP traffic.**

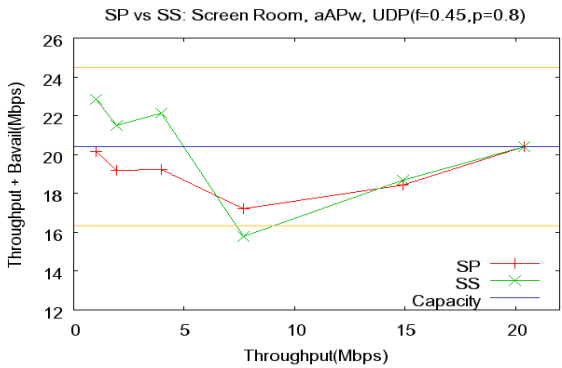

**Figure 7 (One-hop): Screen Room One-hop throughput comparison SP vs SS**

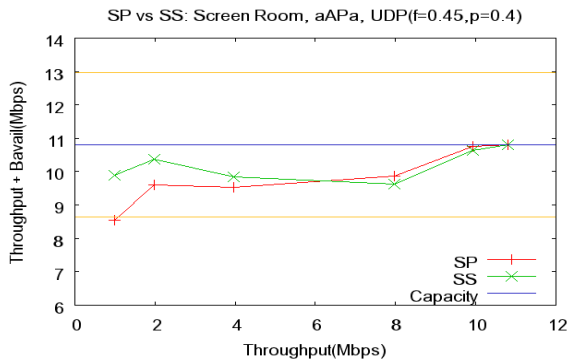

**Figure 8 (Two-hop): Screen Room Two-hop throughput comparison SP vs SS**

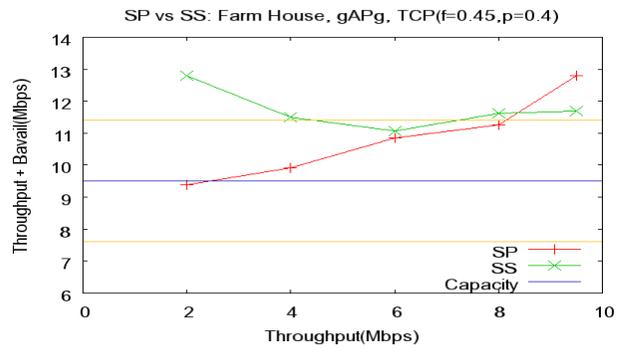

**Figure 9 (TCP): Farm House Two-hop TCP throughput comparison SP vs SS**

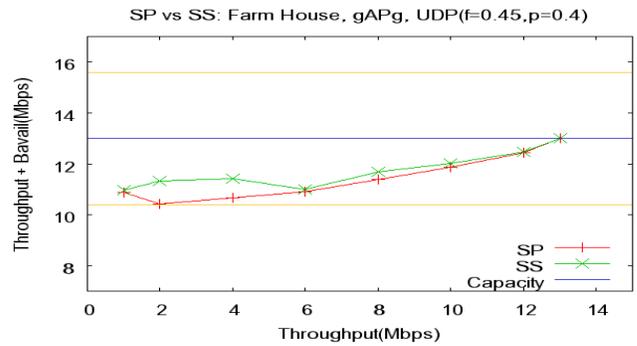

**Figure 10 (UDP): Farm House Two-hop UDP throughput comparison SP vs SS**

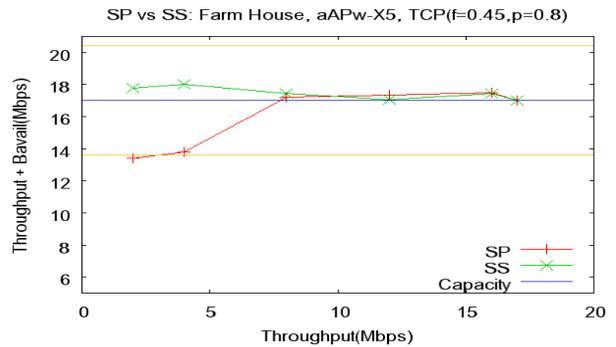

**Figure 11 (TCP): Farm House One-hop throughput SP vs SS with 5Mbps cross-traffic**

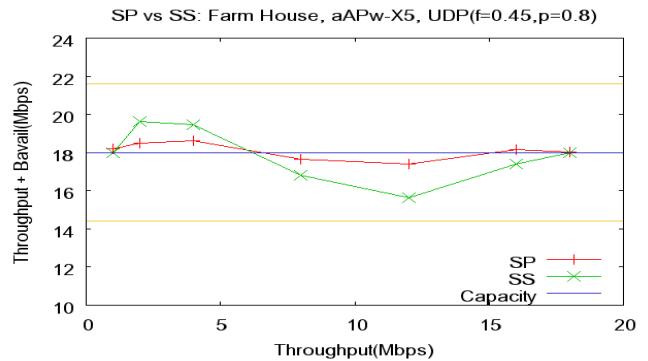

**Figure 12 (UDP): Farm House One-hop throughput SP vs SS with 5Mbps cross-traffic**